\pdfoutput=1
\documentclass[%
 amsmath,amssymb,
twocolumn,
]{revtex4-1}
\usepackage[usenames]{color}
\usepackage{ctable}
\usepackage{graphicx}
\usepackage{dcolumn}
\usepackage{bm}
\usepackage[mathlines]{lineno}
\usepackage{rotating}

\begin{document}


\title{Experimental evidence of a state-point dependent scaling exponent of liquid dynamics}

\author{Alejandro Sanz} \email{asanz@ruc.dk} \affiliation{Glass and
  Time, IMFUFA, Department of Science and Environment, Roskilde
  University, Postbox 260, DK-4000 Roskilde,
  Denmark}

\author{Tina Hecksher} \author{Henriette Wase Hansen} \author{Kristine
  Niss} \author{Ulf R. Pedersen} \email{urp@ruc.dk} \affiliation{Glass
  and Time, IMFUFA, Department of Science and Environment, Roskilde
  University, Postbox 260, DK-4000 Roskilde, Denmark}%

\date{\today}

\begin{abstract}
  A large class of liquids have hidden scale invariance characterized
  by a scaling exponent.  In this letter we present experimental
  evidence that the scaling exponent of liquid dynamics is state-point
  dependent for the glass-forming silicone oil
  tetramethyl-tetraphenyl-trisiloxane (DC704) and 5-polyphenyl ether
  (5PPE).  From dynamic and thermodynamic properties at equilibrium,
  we use a method to estimate the value of $\gamma$ at any state point
  of the pressure-temperature plane, both in the supercooled and
  normal liquid regimes. We find agreement between the average
  exponents and the value obtained by superposition of relaxation
  times over a large range of state-points. We confirm the state-point
  dependence of $\gamma$ by reanalyzing data of 20 metallic liquids
  and two model liquids.
\end{abstract}

\pacs{Valid PACS appear here}
\maketitle


Decreasing the temperature ($T$) or increasing the density ($\rho$) by applying pressure of
liquids lead to slowing down of the molecular dynamics and eventually
a glass transition if crystallization is avoided
\cite{bin05,Simionesco}. It has been demonstrated that for numerous
low molecular weight liquids and polymers, the relaxation time or
other dynamic quantities can be superimposed onto a master curve
within the experimental uncertainty when plotted as a function of
$\rho^{\gamma}/T$. The scaling exponent $\gamma$ is thus sometimes
referred to as a material constant \cite{Tolle,Casalini,Casalini2}. In
this letter, we show that $\gamma$ is in fact state-point dependent
and can be measured at a single state point. In the following we will
use subscripts to distinguish definitions of $\gamma$'s.

Let $\tau$ be the structural relaxation time measured from the
loss-peak frequency of the electric permittivity. We will express
$\tau$ in reduced units of $\sqrt{m/k_BT\rho^{2/3}}$ where $m$ is a
atomic/molecular/polymer-segment mass and define the scaling exponent
of $\tau$ as \cite{gna09}
\begin{equation}
  \gamma_{\tau}(\rho,T) \equiv \left(\frac{\partial\log
      T}{\partial\log\rho}\right)_\tau. 
\label{eq:Eq1}
\end{equation}
In this general definition the exponent depends on the state point and
the fixed quantity ($\tau$ in the above case). A physical
interpretation of $\gamma_\tau$ is that it quantifies the relative
contribution of volume and thermal energy to the temperature evolution
of the molecular mobility \cite{nis07}.
%
Similar to the structural relaxation time, configurational adiabats
can also be associated with a scaling exponent:
\begin{equation}
  \gamma_{S_\textrm{ex}}(\rho,T)\equiv\left(\frac{\partial\log
      T}{\partial\log\rho}\right)_{S_\textrm{ex}} 
\label{eq:Eq12}.
\end{equation}
In general, two scaling exponents of different observables
(structural, dynamical or thermodynamic) will have different
values. However, if the system has so-called ``hidden
scale-invariance'' \cite{gna09,sch14} then all scaling exponents of
different properties will have the same value. In Ref.\ \cite{Ditte}
it was experimentally shown that
$\gamma_\tau = \gamma_{S_\textrm{ex}}$ at one state point of the
silicone oil DC704. This result is spectacular since hidden
scale-invariance can only be present in a class of systems \cite{Ulf,scl_II}. This class
is believed to include systems where van der Waals (vdW) interactions
dominate, but exclude systems where hydrogen-bondes (HB) dominates the
Hamiltonian \cite{Dyre2}.

The isomorph theory \cite{Ulf,gna09,Dyre2,sch14} is a framework for
describing systems where the potential energy function $U(\bf R)$
possesses hidden scale invariance (here, $\bf R$ is the collective
coordinate of the system). Formally hidden scale invariance can be
formulated as the following criterion for two configurations $a$
and $b$: if $U({\bf R}_a)<U({\bf R}_b)$ then
$U(\lambda{\bf R}_a)<U(\lambda{\bf R}_b)$ to a good approximation \cite{sch14}. For
these systems, scaling exponents of many properties have the same
state-point dependence. Thus, there is only one scaling exponent that
give the slope of the so-called isomorphs \cite{gna09} along which dynamical,
structural and many thermodynamic properties are constant in reduced
units. As an example of applications, the framework of the isomorph theory explains
Rosenfeld's excess entropy scaling law \cite{ros77,mit06,ros99,jak15},
i.e. that relaxation time is a function of entropy: $\tau(S_{ex})$, by
equating Eqs. \ref{eq:Eq1} and \ref{eq:Eq12}, and recently the framework has been used to make predictions for
properties of the melting line \cite{ped18}. For the latter, the state point dependence of the scaling exponent is an essential ingredient.

As a special case, Alba-Simionesco, Kivelson and Tarjus (AKT) have
investigated the validity of a scaling law for activated dynamics
where the scaling exponent only depends on density
\cite{alb02,alb04,alb06}, but not on temperature.  Regarding $\gamma$
a material constant is an even more constraining assumption, which is
only valid if the potential part of the Hamiltonian can be
approximated by a sum of inverse power-laws (IPL) $r^{-n}$ pair interactions plus an
arbitrary constant (the IPL hypothesis) \cite{hoo71,Koperwas,pap09,pal10,Romanini,Tolle}. Then the scaling exponent is
independent of state-point ($\gamma_\textrm{IPL}=n/3$) and the relaxation time falls on a master curve when plottet along $\rho^\gamma/T$. In this letter we
wish to investigate the state point dependence of $\gamma$ without any
assumptions and emphasize error estimates. This information can be
used to determine whether: a) $\gamma$ is constant (the IPL hypothesis), b)
$\gamma$ is only a function of $\rho$ (the AKT hypothesis), or c) $\gamma$ is a function of
two thermodynamical variables (the generalized framework of the isomorph theory).

To this aim, we first give an expression for $\gamma_{\tau}$ in terms
of quantities that can be measured at isobaric or isothermal
conditions where most experiments are performed. We define a
generalized fragility
\cite{ang85,Angell2,alb06,nis07,tar14,Casalini2}, inspired by Angell's
suggestion \cite{ang85}:
\begin{equation}
  m_B^A \equiv \left(\frac{\partial\log\tau}{\partial  A}\right)_B,
\label{eq:Eq2}
\end{equation}
where $A$ and $B$ are thermodynamic variables such as $T$, $\rho$ or
pressure $p$. We note that the temperature-fragility $m^T_p$ (which is negative) is
related to the apparent activation enthalpy \cite{Casalini2}
$H_p\equiv k_B[\partial\log\tau/\partial(1/T)]_p=-k_BT^2m^T_p$, and
when the temperature-fragility $m^T_p$ is evaluated at the glass
transition temperature $T_g$ at ambient pressure (0.1 MPa) it is
related to the dimensionless fragility index originally proposed by
Angell \cite{Angell}:
$m_\textrm{Angell}\equiv[\partial\log\tau/\partial (T_g/T)
]_{p,T=T_g}=-T_gm^{T=T_g}_{p=0.1\text{ MPa}}$. The isochoric
activation energy $E_V$ \cite{Casalini2} can also be expressed via the
generalized fragility:
$E_V\equiv k_B[\partial \log\tau/\partial(1/T)]_\rho=-k_BT^2m_\rho^T$.
\ Note that we have {\emph{not}} defined the generalized fragility as
a dimensionless Angell-type index. 

Below we consider the pressure-fragility $m^p_T$ and
temperature-fragility $m^T_p$ as the quantities that are directly
experimentally accessible.  
We rewrite $\gamma_{\tau}$, Eq.\ \ref{eq:Eq1}, in terms of the ratio between two
generalized fragilities by using the thermodynamic identity
$ \left(\frac{\partial \log T}{\partial \log \rho}\right)_\tau = -
\left(\frac{\partial \log\tau}{\partial \log\rho}\right)_T
\left(\frac{\partial \log T}{\partial \log\tau}\right)_\rho,
\label{eq:Eq4} 
$:
\begin{equation}
\gamma_{\tau}=-\frac{\rho m^\rho_T}{T m^T_\rho}\,.
\label{eq:Eq5}  
\end{equation} 
However, typically the pressure, and not the density is controlled
in an experiment. Thus, we wish to have an expression involving
$m^p_T$ and $m^T_p$. From the chain rule
$ m^\rho_T = \left(\frac{\partial\log\tau}{\partial
    p}\right)_T\left(\frac{\partial p}{\partial\rho}\right)_T, $ it
follows that
\begin{equation}
m^\rho_T = K_{T}m^p_T/\rho\,,
\end{equation}
where $K_{T}=\left(\frac{\partial p}{\partial \log\rho}\right)_T$ is
the isothermal bulk modulus. Using the identity
$ \left(\frac{\partial \log\tau}{\partial T}\right)_\rho
=\left(\frac{\partial \log\tau}{\partial T}\right)_p +
\left(\frac{\partial \log\tau}{\partial p}\right)_T
\left(\frac{\partial p}{\partial T}\right)_\rho,
\label{eq:Eq6}  
$ and inserting the thermal-expansion coefficient at constant pressure
$\alpha_{p}=-\left(\frac{\partial \log\rho}{\partial T}\right)_p$, we
obtain
\begin{equation}
m_\rho^T = m_p^T + \alpha_p K_T m_T^p\,.
\label{eq:Eq7} 
\end{equation}
Finally, by combining the three numbered equations above we arrive at an expression that
relates $\gamma_{\tau}$ to directly measurable properties at
equilibrium without needing the superposition of relaxation times:
\begin{equation}
\gamma_{\tau} = -\frac{K_T m_T^p}{T m^T_p+\alpha_{p}TK_Tm_T^p}.
\label{eq:Eq8} 
\end{equation}
Thus, the state point dependence of the scaling exponent
$\gamma_{\tau}$ can be obtained from the thermal-expansion coefficient
($\alpha_{p}$), pressure and temperature fragilities ($m_T^p$ and
$m_p^T$) and isothermal bulk modulus ($K_T$). The idea of computing
$\gamma$ from two fragilities has been
applied in other studies \cite{alb02,nis07,Casalini2,tar14}. In this
study we suggest to generalize this approach to the entire liquid
state away from the glass transition. Due to the measuring methods
favored in our lab \cite{Tage}, we choose to get the value of the
isothermal bulk modulus from the adiabatic modulus $K_S$ measured by
the speed of sound \cite{Tage}: $
K_T=K_S \rho C_p/[\rho C_p+T \alpha_{p}^2 K_S]
$
where $C_p$ is the isobaric heat capacity measured with differential
scanning calorimetry.
%


\begin{figure}
  \centering
  \includegraphics[trim = 15mm 60mm 15mm 50mm, clip,
  width=0.85\linewidth]{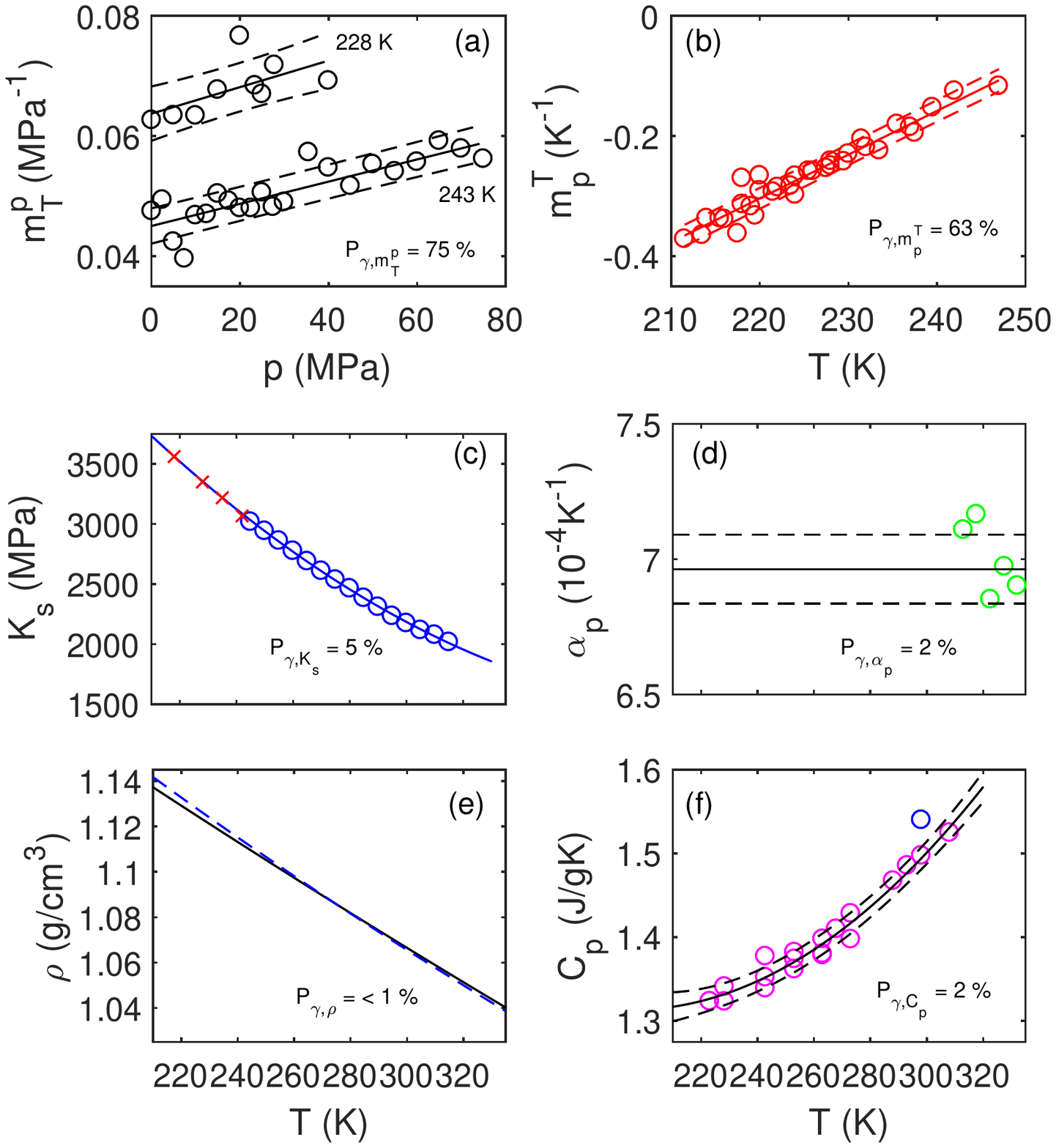}
  \caption{Experimental dynamic and thermodynamic properties measured
    by our group and used for calculating the exponent $\gamma_{\tau}$
    at two different state points, (228 K, 0.1 MPa) and (243 K, 0.1
    MPa), for the silicone oil DC704. (a) Pressure fragility at two
    isotherms as a function of pressure from dielectric relaxation
    measurements. (b) Temperature fragility at 0.1 MPa as a function
    of temperature from dielectric relaxation measurements. (c)
    Adiabatic bulk modulus from the standing waves in a bulk
    transducer. Red crosses highlight the values of $K_s$ used for
    estimated $\gamma_{\tau}$ at four different temperatures. (d)
    Thermal expansion coefficient extracted from PVT measurements. (e)
    Temperature evolution of density for liquid DC704 from the Tait
    equation (solid line) and from the extrapolation of the values at
    room conditions by using the expansion coefficient (dashed
    line). (f) Heat capacity as a function of temperature for liquid
    DC704 from DSC measurements. Blue circles included in panel (f)
    corresponds to literature data \cite{Klein}. The $P_\gamma,_i$
    values in each panel indicate the contributions to the statistical
    error on the estimate of $\gamma_\tau$ at $T=288$ K.}
	\label{fig:Fig1}
\end{figure}

\begin{figure}
  \centering
  \includegraphics[trim = 0mm 30mm 0mm 30mm, clip,
  width=0.85\linewidth]{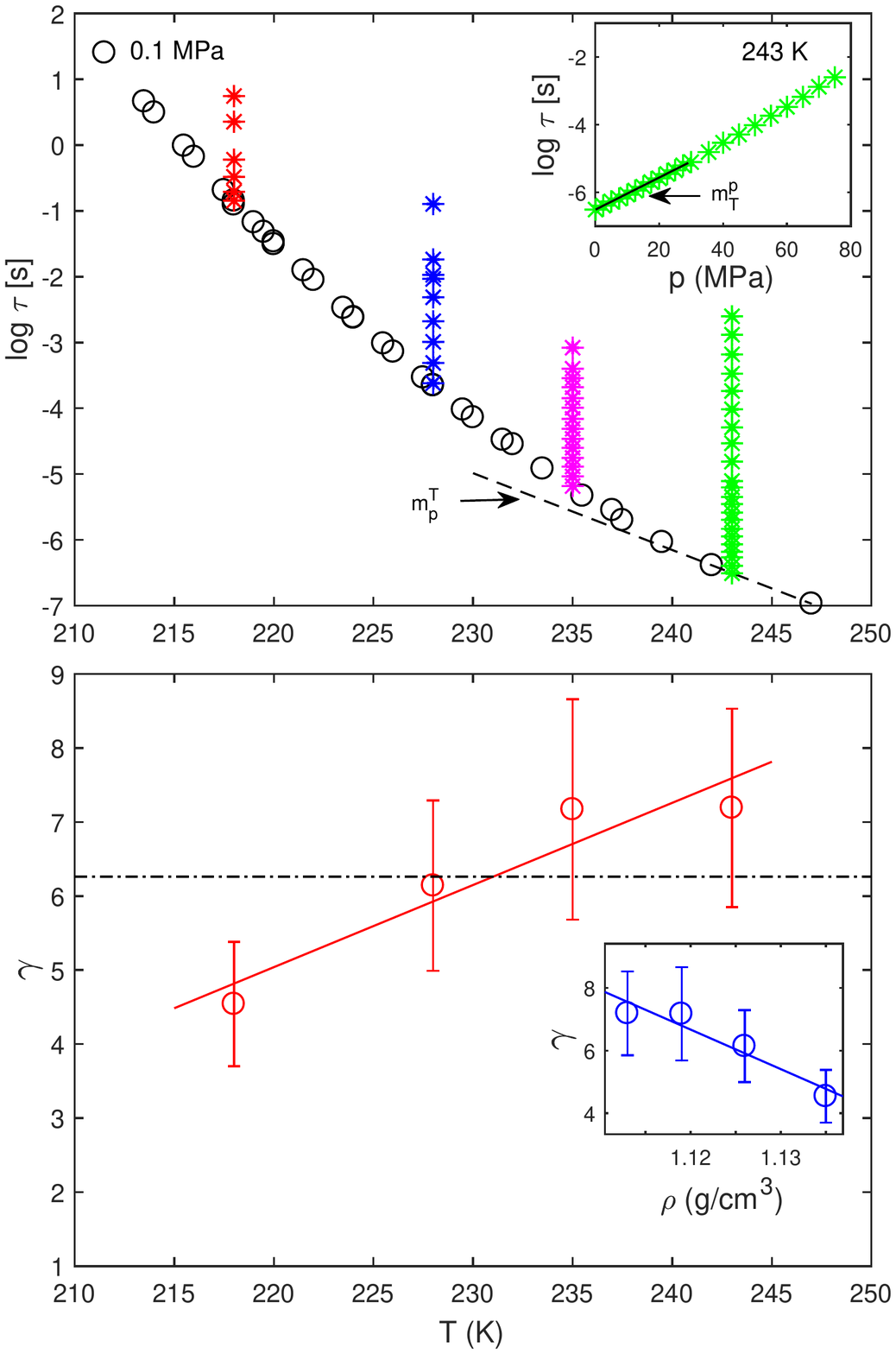}
  \caption{(a) The structural relaxation time $\tau$ of DC704 measured
    by the dielectric loss. (b) Density scaling exponent
    (Eq. \ref{eq:Eq8}) as a function of temperature at atmospheric
    pressure (0.1 MPa). Black line indicates the average value of
    $\gamma_{\tau}$ and the red line is a guide to the eye. The inset
    displays $\gamma_{\tau}$ as a function of density. Blue line is a
    guide to the eye. }
	\label{fig:Fig3}
\end{figure}

We focus our investigation to two well-studied van der Waals liquids,
tetramethyl-tetraphenyl-trisiloxane (DC704) and 5-polyphenyl ether
(5PPE), with values of $T_{g}$ at atmospheric pressure of 211 and 245
K respectively \cite{Lisa}. As an example, in Fig. \ref{fig:Fig1} we
collect the quantities we need to calculate $\gamma_{\tau}$ by using
Eq. \ref{eq:Eq8} where $K_T$ is computed from $K_s$ and $C_p$ as mentioned above. 
%
Figure \ref{fig:Fig3}(a) show the relaxation times at atmospheric pressure for the silicone oil DC704, and Fig.\ \ref{fig:Fig3}(b) displays $\gamma_{\tau}$ at four temperatures along the 0.1 MPa isobar.

To evaluate to what extent the measured quantities contribute to the
final error of $\gamma_{\tau}$ and, in this way, to predict where one
should give particular attention to reduce as much as possible the
uncertainty of the data, we use statistical tools for analyzing the
results. A large population of values for each variable ($N = 10^{5}$)
are sampled by using a Monte Carlo approach, assuming a normal
distribution centered about its mean within an interval determined by
the corresponding standard deviation. The error is calculated by
sampling a random collection of different scenarios for the variables
in Eq. \ref{eq:Eq8}.
The $P_\gamma,_i$ values in the panels of Fig. \ref{fig:Fig1}
represent the pairwise Pearson correlation coefficients between
$\gamma_{\tau}$ and the variables involved in the computation of
Eq. \ref{eq:Eq8}. If $P_\gamma,_i$ equals 100\%, there is
a total positive correlation and 0 would indicate absolute lack of
correlation. The property that has the strongest correlation with
$\gamma_{\tau}$ is the pressure fragility, $m_T^p$, followed by the
temperature fragility, $m_p^T$. It is therefore recommended to
measure the generalized fragilities with high accuracy in order to
reduce the uncertainty in $\gamma_{\tau}$. On the contrary, slight
variations in the density $\rho$ and in the thermal-expansion
coefficient $\alpha$ has a minor effect on the resulting values of
$\gamma_{\tau}$. The final error estimate on $\gamma_{\tau}$ are shown as errorbars on Fig.\ \ref{fig:Fig3}(b) suggesting an increase of $\gamma_{\tau}$ as temperature increases or the
density is decreases.

To challenge the robustness of our approach, we collect in Table
\ref{tab:table1} the values of $\gamma_{\tau}$ for other systems,
including two hydrogen-bonded liquids. We also compare our results
with other procedures, such as the superposition of relaxation times
($\gamma_\textrm{IPL}$), the expression proposed by Casalini and
Roland ($\gamma_*$), and an experimental prediction of the isomorph
theory ($\gamma_\textrm{isom}$) mentioned in the introduction. The
route to the scaling exponent proposed by Casalini and Roland, in which
continuity of the entropy at the glass transition is assumed, has been
shown to work effectively in several van der Waals and hydrogen-bonded
glass-forming liquids, as well as in polymeric materials
\cite{Casalini2,Romanini}. From Table \ref{tab:table1}, one observes
that the values of $\gamma$'s obtained through these three methods are
in fair agreement with the average value we would obtain using
Eq. \ref{eq:Eq8} from the results at different state points. Table
\ref{tab:table1} includes information on the class of liquid and the
method utilized for obtaining the bulk modulus.

\begin{table*}
  \caption{\label{tab:table1}%
    Density-scaling exponent $\gamma_{\tau}$ at $p=0.1$ MPa for
    selected substances obtained by using Eq. \ref{eq:Eq8} and its
    comparison with other methods.}
  \begin{ruledtabular}
    \begin{tabular}{ccccccccccc}
      \textrm{Substance}& T & \textrm{$T_{g}$(0.1
                              MPa)}&\textrm{$\gamma_{\tau}$}\footnote{Scaling
                                     exponent
                                                         computed in
                                                         this study
                                                         via
                                                         Eq. \ref{eq:Eq8}.}&
                                                                             \textrm{$\gamma_\textrm{IPL}$}\footnote{Scaling exponent computed with superposition of relaxation times (assuming the IPL hypothesis for pair interactions).} &
                                                                                                                                                                                                                                                                                                                           \textrm{$\gamma_*$ }\footnote{A state-point independent scaling exponent calculated by Casalini et al.\ from static ambient-pressure quantities.}&\textrm{$\gamma_\textrm{isom}$}\footnote{Scaling exponent calculated from static properties using an expression derived from isomorph theory (DC704 at 214 K, 0.1 MPa).}&\textrm{Class}& \textrm{References}
      \\
      \colrule
      DC704 & 218 K& 211 K & 4.5 $\pm$ 0.8 & 6.2 $\pm$ 0.2 & 6.8 $\pm$ 0.8 & 6.0 $\pm$ 2.0  &vdW&\cite{Ditte,Casalini3,Casalini2}\\
      -      & 228 K& - & 6.1 $\pm$ 1.1 & - & - & -  &vdW& \\
      -     & 235 K & - & 7.1 $\pm$ 1.5 & - & - & -  &vdW&\\
      -     & 242 K & - &  7.2 $\pm$ 1.3 & - & - & -  &vdW&\\
      5PPE  &268 K & 245 K & 5.4 $\pm$ 0.2 & 5.5 $\pm$ 0.3 & -& - &vdW&\cite{Ditte2,Adrjanowicz}\\
      -    & 284 K & - & 7.8 $\pm$ 0.5 & - & -& -   &vdW&\\
      Glycerol & 230 K& 185 K & 0.88 $\pm$ 0.09 & 1-1.8  & 1.28 $\pm$ 0.15& - &HB&\cite{Dreyfus,Casalini3,Casalini2,Klieber,Roland}\\
      -      &258 K & - & 0.43 $\pm$ 0.05 & -  & -& -  &HB&\\
      DPG &240 K & 195 K & 1.3 $\pm$ 0.2 & 1.5-1.99  & -& -  &HB&\cite{Grzybowski2,Preparation}\\			
		\end{tabular}
	\end{ruledtabular}
\end{table*}

\begin{figure}
  \centering
  \includegraphics[trim = 0mm 5mm 20mm 12mm, clip,
  width=0.45\linewidth]{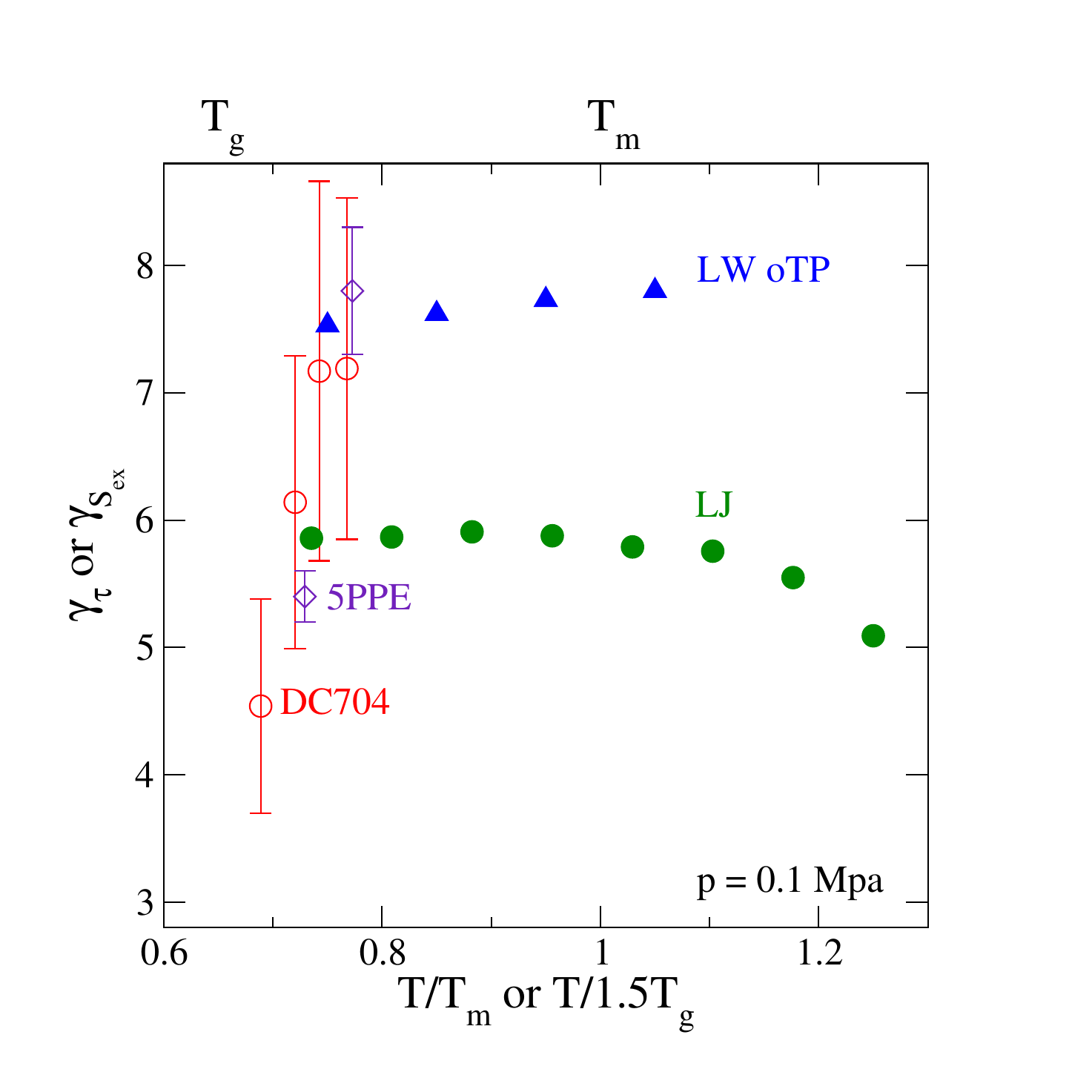}
  \includegraphics[trim = 5mm 5mm 20mm 20mm, clip,
  width=0.45\linewidth]{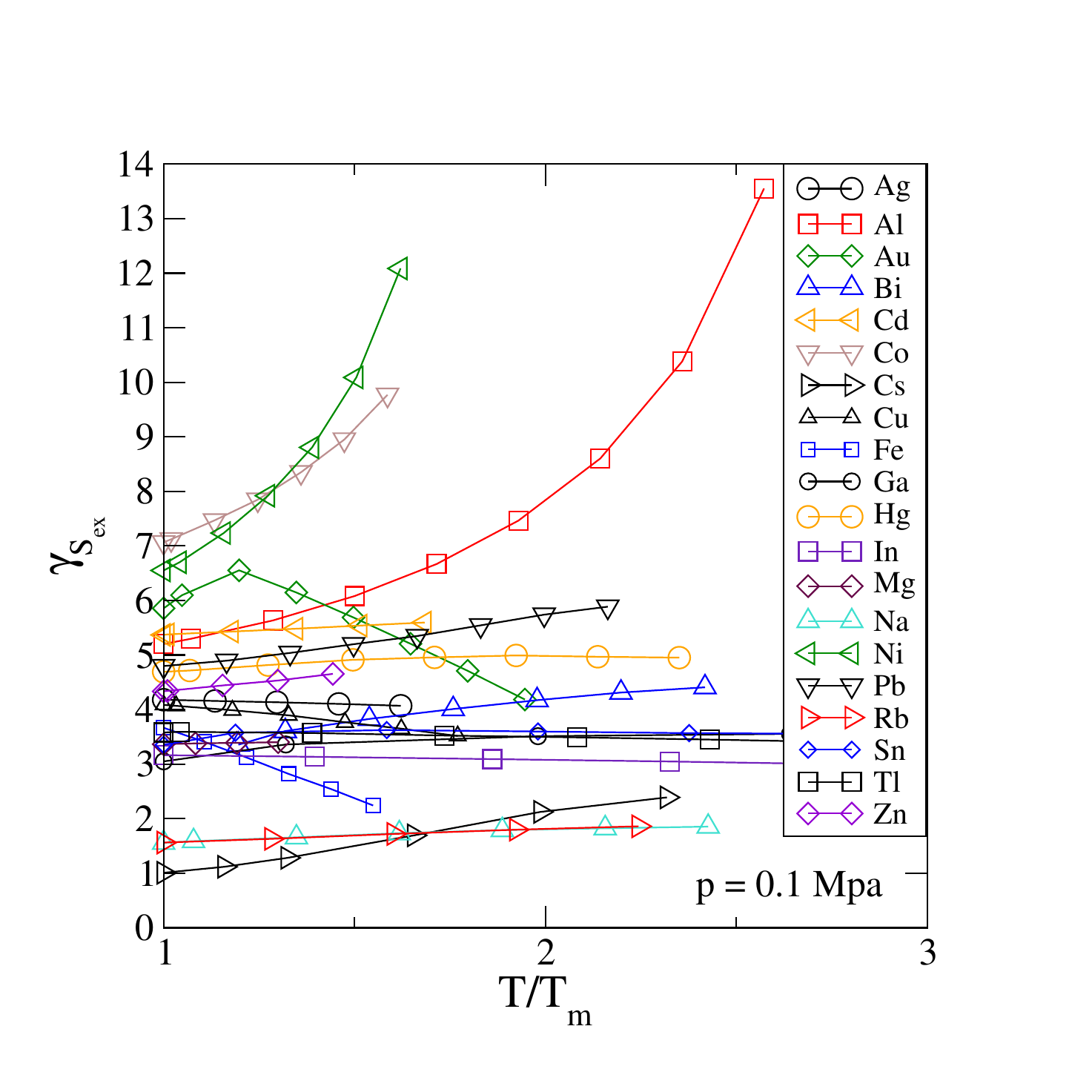}
  \caption{(a) Comparison of experimentally determined scaling
    exponents (open symbols) with values of two model liquids
    (filled symbols). (b) The scaling exponent of the configurational
    adiabat $\gamma_{S_\textrm{ex}}$ for liquid metals.}
  \label{fig:Fig4}
\end{figure}

Figure \ref{fig:Fig4} shows that the scaling exponents is also
state-point dependent for model liquids and metals. Figure
\ref{fig:Fig4}(a) shows $\gamma_{S_{ex}}$ computed from molecular
dynamics simulations \cite{rumd} of the Lennard-Jones (LJ) liquid
\cite{lj24} and a Lennard-Jones trimer suggested by Lewis and
Wahnstr{\"o}m \cite{lew93} as a coarse-grained model for
ortho-terphenyl (LW-oTP). In line with the experimental findings, the
LW-oTP model show an increase of the exponent with temperature, though
less dramatic. In Fig.\ \ref{fig:Fig4}(b) we reanalyze experimental
data for 20 metallic liquids \cite{sin07}, including metals where {\it
  ab initio} density functional theory calculations show hidden-scale
invariance \cite{hum15}. For the mono atomic metallic liquids, the scaling exponent of excess entropy
is estimated using
$\gamma_{S_{ex}}=[\gamma_G-k_B/c_v]/[1-3k_B/c_v]$, where
$\gamma_G=\alpha_pK_T/\rho c_v$ is the thermodynamic Gr{\"u}neisen
parameter \cite{gru12} (it is assumed that the material is above the Debye
temperature \cite{scl_II,hum15}). The exponents have significant
temperature dependencies with both positive and negative slopes along
the $p=0.1$~MPa isobar.

These results are in line with the state point dependence of the scaling
exponent $\gamma$ by simulations of Kob-Andersen binary
Lennard-Jones liquids \cite{Bohling}, and also consistent with the
generalized scaling equation of state reported in
ref. \cite{Grzybowski}. The latter study proposes a density dependence
function for the scaling exponent with two parameters that can be
estimated from their generalized density-scaling equation of state
\cite{Grzybowski}. In this way, it is possible to determine the
evolution of $\gamma$ with density from PVT measurements as it was
recently reported for several organic liquids \cite{Lopez}.

An advantage of Eq.\ \ref{eq:Eq8} is that it allow the determination of $\gamma_{\tau}$ for any
kind of liquid regardless of its vitrification ability, and the
formalism is not assuming activated dynamics \cite{alb02}. Thus, we
suggest a procedure that can potentially be used at any given point in
the temperature-pressure plane. We note that the average
$\gamma_{\tau}$ found here ($6.1\pm1.1$ for the investigated state
points of the silicone oil DC704) is in good agreement with the
exponent estimated by other methods: i) the results obtained by
superposing relaxation times \cite{Ditte}, ii) the experimental
prediction of the isomorph theory \cite{Ditte}, and iii) the value
obtained through the equation derived by Casalini and Roland
\cite{Casalini2}, confirming the validity of Eq. \ref{eq:Eq8} for the
universal prediction of $\gamma_{\tau}$.
In experimental studies, changes in the density are usually small, so
in the scaling relation, $X = f (\rho^{\gamma_{\textrm{IPL}}} / T)$,
the exponent is generally assumed to be state-point independent, with
$X$ a measure of the molecular mobility and $f$ a function that is a
priori unknown. Nonetheless, within the framework of the isomorph
theory, the exponent may depend on the state point as it was
corroborated by computer simulations of strongly correlating liquids
(see Fig.\ref{fig:Fig4}), that is, those exhibiting hidden scale
invariance such as van der Waals molecules and metals
\cite{Bohling,Dyre2}.


In summary, we have shown that it is possible to determine the scaling
parameter $\gamma_{\tau}$ from dynamic and thermodynamic properties of
liquids at a single state-point, and that $\gamma_{\tau}$ can be state
point dependent. The expression presented in Eq.\ \ref{eq:Eq8}
connects $\gamma_{\tau}$ to measurable quantities that can be
potentially estimated in a wide range of thermodynamic conditions,
from state points near the glass transition to well above the melting
point (including elevated pressures). This new route is free of
assumptions and can be utilized for both glass- and non-glass-forming
materials.

\section{ACKNOWLEDGMENTS}
This study was initiated after a series of discussion between Niels Boye Olsen and URP.
We thank Jeppe C. Dyre for valuable input to the manuscript.
This work is supported by the VILLUM Foundation’s
VKR-023455 and Matter (16515) grants.

\bibliography{Biblio_gamma}

\begin{thebibliography}{49}%
\makeatletter
\providecommand \@ifxundefined [1]{%
 \@ifx{#1\undefined}
}%
\providecommand \@ifnum [1]{%
 \ifnum #1\expandafter \@firstoftwo
 \else \expandafter \@secondoftwo
 \fi
}%
\providecommand \@ifx [1]{%
 \ifx #1\expandafter \@firstoftwo
 \else \expandafter \@secondoftwo
 \fi
}%
\providecommand \natexlab [1]{#1}%
\providecommand \enquote  [1]{``#1''}%
\providecommand \bibnamefont  [1]{#1}%
\providecommand \bibfnamefont [1]{#1}%
\providecommand \citenamefont [1]{#1}%
\providecommand \href@noop [0]{\@secondoftwo}%
\providecommand \href [0]{\begingroup \@sanitize@url \@href}%
\providecommand \@href[1]{\@@startlink{#1}\@@href}%
\providecommand \@@href[1]{\endgroup#1\@@endlink}%
\providecommand \@sanitize@url [0]{\catcode `\\12\catcode `\$12\catcode
  `\&12\catcode `\#12\catcode `\^12\catcode `\_12\catcode `\%12\relax}%
\providecommand \@@startlink[1]{}%
\providecommand \@@endlink[0]{}%
\providecommand \url  [0]{\begingroup\@sanitize@url \@url }%
\providecommand \@url [1]{\endgroup\@href {#1}{\urlprefix }}%
\providecommand \urlprefix  [0]{URL }%
\providecommand \Eprint [0]{\href }%
\providecommand \doibase [0]{http://dx.doi.org/}%
\providecommand \selectlanguage [0]{\@gobble}%
\providecommand \bibinfo  [0]{\@secondoftwo}%
\providecommand \bibfield  [0]{\@secondoftwo}%
\providecommand \translation [1]{[#1]}%
\providecommand \BibitemOpen [0]{}%
\providecommand \bibitemStop [0]{}%
\providecommand \bibitemNoStop [0]{.\EOS\space}%
\providecommand \EOS [0]{\spacefactor3000\relax}%
\providecommand \BibitemShut  [1]{\csname bibitem#1\endcsname}%
\let\auto@bib@innerbib\@empty
\bibitem [{\citenamefont {Binder}\ and\ \citenamefont {Kob}(2005)}]{bin05}%
  \BibitemOpen
  \bibfield  {author} {\bibinfo {author} {\bibfnamefont {K.}~\bibnamefont
  {Binder}}\ and\ \bibinfo {author} {\bibfnamefont {W.}~\bibnamefont {Kob}},\
  }\href@noop {} {\emph {\bibinfo {title} {Glassy Materials and Disordered
  Solids: An Introduction to their Statistical Mechanics}}}\ (\bibinfo
  {publisher} {World Scientific, Singapore},\ \bibinfo {year}
  {2005})\BibitemShut {NoStop}%
\bibitem [{\citenamefont {Alba-Simionesco}\ \emph
  {et~al.}(2004{\natexlab{a}})\citenamefont {Alba-Simionesco}, \citenamefont
  {Cailliaux}, \citenamefont {Alegria},\ and\ \citenamefont
  {Tarjus}}]{Simionesco}%
  \BibitemOpen
  \bibfield  {author} {\bibinfo {author} {\bibfnamefont {C.}~\bibnamefont
  {Alba-Simionesco}}, \bibinfo {author} {\bibfnamefont {A.}~\bibnamefont
  {Cailliaux}}, \bibinfo {author} {\bibfnamefont {A.}~\bibnamefont {Alegria}},
  \ and\ \bibinfo {author} {\bibfnamefont {G.}~\bibnamefont {Tarjus}},\ }\href
  {\doibase 10.1209/epl/i2004-10214-6} {\bibfield  {journal} {\bibinfo
  {journal} {Europhysics Letters}\ }\textbf {\bibinfo {volume} {68}},\ \bibinfo
  {pages} {58} (\bibinfo {year} {2004}{\natexlab{a}})}\BibitemShut {NoStop}%
\bibitem [{\citenamefont {T\"olle}\ \emph {et~al.}(1998)\citenamefont
  {T\"olle}, \citenamefont {Schober}, \citenamefont {Wuttke}, \citenamefont
  {Randl},\ and\ \citenamefont {Fujara}}]{Tolle}%
  \BibitemOpen
  \bibfield  {author} {\bibinfo {author} {\bibfnamefont {A.}~\bibnamefont
  {T\"olle}}, \bibinfo {author} {\bibfnamefont {H.}~\bibnamefont {Schober}},
  \bibinfo {author} {\bibfnamefont {J.}~\bibnamefont {Wuttke}}, \bibinfo
  {author} {\bibfnamefont {O.~G.}\ \bibnamefont {Randl}}, \ and\ \bibinfo
  {author} {\bibfnamefont {F.}~\bibnamefont {Fujara}},\ }\href {\doibase
  10.1103/PhysRevLett.80.2374} {\bibfield  {journal} {\bibinfo  {journal}
  {Phys. Rev. Lett.}\ }\textbf {\bibinfo {volume} {80}},\ \bibinfo {pages}
  {2374} (\bibinfo {year} {1998})}\BibitemShut {NoStop}%
\bibitem [{\citenamefont {Casalini}\ and\ \citenamefont
  {Roland}(2004)}]{Casalini}%
  \BibitemOpen
  \bibfield  {author} {\bibinfo {author} {\bibfnamefont {R.}~\bibnamefont
  {Casalini}}\ and\ \bibinfo {author} {\bibfnamefont {C.}~\bibnamefont
  {Roland}},\ }\href {\doibase 10.1103/PhysRevE.69.062501} {\bibfield
  {journal} {\bibinfo  {journal} {Physical Review E}\ }\textbf {\bibinfo
  {volume} {69}},\ \bibinfo {pages} {062501} (\bibinfo {year}
  {2004})}\BibitemShut {NoStop}%
\bibitem [{\citenamefont {Casalini}\ and\ \citenamefont
  {Roland}(2014)}]{Casalini2}%
  \BibitemOpen
  \bibfield  {author} {\bibinfo {author} {\bibfnamefont {R.}~\bibnamefont
  {Casalini}}\ and\ \bibinfo {author} {\bibfnamefont {C.~M.}\ \bibnamefont
  {Roland}},\ }\href {\doibase 10.1103/PhysRevLett.113.085701} {\bibfield
  {journal} {\bibinfo  {journal} {Physical Review Letters}\ }\textbf {\bibinfo
  {volume} {113}},\ \bibinfo {pages} {085701} (\bibinfo {year}
  {2014})}\BibitemShut {NoStop}%
\bibitem [{\citenamefont {Gnan}\ \emph {et~al.}(2009)\citenamefont {Gnan},
  \citenamefont {Schr{\o}der}, \citenamefont {Pedersen}, \citenamefont
  {Bailey},\ and\ \citenamefont {Dyre}}]{gna09}%
  \BibitemOpen
  \bibfield  {author} {\bibinfo {author} {\bibfnamefont {N.}~\bibnamefont
  {Gnan}}, \bibinfo {author} {\bibfnamefont {T.~B.}\ \bibnamefont
  {Schr{\o}der}}, \bibinfo {author} {\bibfnamefont {U.~R.}\ \bibnamefont
  {Pedersen}}, \bibinfo {author} {\bibfnamefont {N.~P.}\ \bibnamefont
  {Bailey}}, \ and\ \bibinfo {author} {\bibfnamefont {J.~C.}\ \bibnamefont
  {Dyre}},\ }\href {\doibase 10.1063/1.3265957} {\bibfield  {journal} {\bibinfo
   {journal} {J. Chem. Phys.}\ }\textbf {\bibinfo {volume} {131}},\ \bibinfo
  {pages} {234504} (\bibinfo {year} {2009})}\BibitemShut {NoStop}%
\bibitem [{\citenamefont {Niss}\ \emph {et~al.}(2007)\citenamefont {Niss},
  \citenamefont {Dalle-Ferrier}, \citenamefont {Tarjus},\ and\ \citenamefont
  {Alba-Simionesco}}]{nis07}%
  \BibitemOpen
  \bibfield  {author} {\bibinfo {author} {\bibfnamefont {K.}~\bibnamefont
  {Niss}}, \bibinfo {author} {\bibfnamefont {C.}~\bibnamefont {Dalle-Ferrier}},
  \bibinfo {author} {\bibfnamefont {G.}~\bibnamefont {Tarjus}}, \ and\ \bibinfo
  {author} {\bibfnamefont {C.}~\bibnamefont {Alba-Simionesco}},\ }\href@noop {}
  {\bibfield  {journal} {\bibinfo  {journal} {J. Phys.: Condens. Matter}\
  }\textbf {\bibinfo {volume} {19}},\ \bibinfo {pages} {076102} (\bibinfo
  {year} {2007})}\BibitemShut {NoStop}%
\bibitem [{\citenamefont {Schr{\o}der}\ and\ \citenamefont
  {Dyre}(2014)}]{sch14}%
  \BibitemOpen
  \bibfield  {author} {\bibinfo {author} {\bibfnamefont {T.~B.}\ \bibnamefont
  {Schr{\o}der}}\ and\ \bibinfo {author} {\bibfnamefont {J.~C.}\ \bibnamefont
  {Dyre}},\ }\href {\doibase http://dx.doi.org/10.1063/1.4901215} {\bibfield
  {journal} {\bibinfo  {journal} {J. Chem. Phys.}\ }\textbf {\bibinfo {volume}
  {141}},\ \bibinfo {pages} {204502} (\bibinfo {year} {2014})}\BibitemShut
  {NoStop}%
\bibitem [{\citenamefont {Gundermann}\ \emph {et~al.}(2011)\citenamefont
  {Gundermann}, \citenamefont {Pedersen}, \citenamefont {Hecksher},
  \citenamefont {Bailey}, \citenamefont {Jakobsen}, \citenamefont
  {Christensen}, \citenamefont {Olsen}, \citenamefont {Schroder}, \citenamefont
  {Fragiadakis}, \citenamefont {Casalini}, \citenamefont {Roland},
  \citenamefont {Dyre},\ and\ \citenamefont {Niss}}]{Ditte}%
  \BibitemOpen
  \bibfield  {author} {\bibinfo {author} {\bibfnamefont {D.}~\bibnamefont
  {Gundermann}}, \bibinfo {author} {\bibfnamefont {U.~R.}\ \bibnamefont
  {Pedersen}}, \bibinfo {author} {\bibfnamefont {T.}~\bibnamefont {Hecksher}},
  \bibinfo {author} {\bibfnamefont {N.~P.}\ \bibnamefont {Bailey}}, \bibinfo
  {author} {\bibfnamefont {B.}~\bibnamefont {Jakobsen}}, \bibinfo {author}
  {\bibfnamefont {T.}~\bibnamefont {Christensen}}, \bibinfo {author}
  {\bibfnamefont {N.~B.}\ \bibnamefont {Olsen}}, \bibinfo {author}
  {\bibfnamefont {T.~B.}\ \bibnamefont {Schroder}}, \bibinfo {author}
  {\bibfnamefont {D.}~\bibnamefont {Fragiadakis}}, \bibinfo {author}
  {\bibfnamefont {R.}~\bibnamefont {Casalini}}, \bibinfo {author}
  {\bibfnamefont {C.~M.}\ \bibnamefont {Roland}}, \bibinfo {author}
  {\bibfnamefont {J.~C.}\ \bibnamefont {Dyre}}, \ and\ \bibinfo {author}
  {\bibfnamefont {K.}~\bibnamefont {Niss}},\ }\href {\doibase
  10.1038/nphys2031} {\bibfield  {journal} {\bibinfo  {journal} {Nature
  Physics}\ }\textbf {\bibinfo {volume} {7}},\ \bibinfo {pages} {816} (\bibinfo
  {year} {2011})}\BibitemShut {NoStop}%
\bibitem [{\citenamefont {Pedersen}\ \emph {et~al.}(2008)\citenamefont
  {Pedersen}, \citenamefont {Bailey}, \citenamefont {Schr{\o}der},\ and\
  \citenamefont {Dyre}}]{Ulf}%
  \BibitemOpen
  \bibfield  {author} {\bibinfo {author} {\bibfnamefont {U.~R.}\ \bibnamefont
  {Pedersen}}, \bibinfo {author} {\bibfnamefont {N.~P.}\ \bibnamefont
  {Bailey}}, \bibinfo {author} {\bibfnamefont {T.~B.}\ \bibnamefont
  {Schr{\o}der}}, \ and\ \bibinfo {author} {\bibfnamefont {J.~C.}\ \bibnamefont
  {Dyre}},\ }\href {\doibase 10.1103/PhysRevLett.100.015701} {\bibfield
  {journal} {\bibinfo  {journal} {Physical Review Letters}\ }\textbf {\bibinfo
  {volume} {100}},\ \bibinfo {pages} {015701} (\bibinfo {year}
  {2008})}\BibitemShut {NoStop}%
\bibitem [{\citenamefont {Bailey}\ \emph {et~al.}(2008)\citenamefont {Bailey},
  \citenamefont {Pedersen}, \citenamefont {Gnan}, \citenamefont {Schrøder},\
  and\ \citenamefont {Dyre}}]{scl_II}%
  \BibitemOpen
  \bibfield  {author} {\bibinfo {author} {\bibfnamefont {N.~P.}\ \bibnamefont
  {Bailey}}, \bibinfo {author} {\bibfnamefont {U.~R.}\ \bibnamefont
  {Pedersen}}, \bibinfo {author} {\bibfnamefont {N.}~\bibnamefont {Gnan}},
  \bibinfo {author} {\bibfnamefont {T.~B.}\ \bibnamefont {Schrøder}}, \ and\
  \bibinfo {author} {\bibfnamefont {J.~C.}\ \bibnamefont {Dyre}},\ }\href
  {\doibase 10.1063/1.2982249} {\bibfield  {journal} {\bibinfo  {journal} {J.
  Chem. Phys.}\ }\textbf {\bibinfo {volume} {129}},\ \bibinfo {pages} {184508}
  (\bibinfo {year} {2008})}\BibitemShut {NoStop}%
\bibitem [{\citenamefont {Dyre}(2014)}]{Dyre2}%
  \BibitemOpen
  \bibfield  {author} {\bibinfo {author} {\bibfnamefont {J.~C.}\ \bibnamefont
  {Dyre}},\ }\href@noop {} {\bibfield  {journal} {\bibinfo  {journal} {The
  Journal of Physical Chemistry B}\ }\textbf {\bibinfo {volume} {118}},\
  \bibinfo {pages} {10007} (\bibinfo {year} {2014})}\BibitemShut {NoStop}%
\bibitem [{\citenamefont {Rosenfeld}(1977)}]{ros77}%
  \BibitemOpen
  \bibfield  {author} {\bibinfo {author} {\bibfnamefont {Y.}~\bibnamefont
  {Rosenfeld}},\ }\href {\doibase 10.1103/PhysRevA.15.2545} {\bibfield
  {journal} {\bibinfo  {journal} {Phys. Rev. A}\ }\textbf {\bibinfo {volume}
  {15}},\ \bibinfo {pages} {2545} (\bibinfo {year} {1977})}\BibitemShut
  {NoStop}%
\bibitem [{\citenamefont {Mittal}\ \emph {et~al.}(2006)\citenamefont {Mittal},
  \citenamefont {Errington},\ and\ \citenamefont {Truskett}}]{mit06}%
  \BibitemOpen
  \bibfield  {author} {\bibinfo {author} {\bibfnamefont {J.}~\bibnamefont
  {Mittal}}, \bibinfo {author} {\bibfnamefont {J.~R.}\ \bibnamefont
  {Errington}}, \ and\ \bibinfo {author} {\bibfnamefont {T.~M.}\ \bibnamefont
  {Truskett}},\ }\href@noop {} {\bibfield  {journal} {\bibinfo  {journal} {J.
  Chem. Phys.}\ } (\bibinfo {year} {2006})}\BibitemShut {NoStop}%
\bibitem [{\citenamefont {Rosenfeld}(1999)}]{ros99}%
  \BibitemOpen
  \bibfield  {author} {\bibinfo {author} {\bibfnamefont {Y.}~\bibnamefont
  {Rosenfeld}},\ }\href {http://stacks.iop.org/0953-8984/11/i=28/a=303}
  {\bibfield  {journal} {\bibinfo  {journal} {Journal of Physics: Condensed
  Matter}\ }\textbf {\bibinfo {volume} {11}},\ \bibinfo {pages} {5415}
  (\bibinfo {year} {1999})}\BibitemShut {NoStop}%
\bibitem [{\citenamefont {Jakse}\ and\ \citenamefont {Pasturel}(2015)}]{jak15}%
  \BibitemOpen
  \bibfield  {author} {\bibinfo {author} {\bibfnamefont {N.}~\bibnamefont
  {Jakse}}\ and\ \bibinfo {author} {\bibfnamefont {A.}~\bibnamefont
  {Pasturel}},\ }\href {\doibase doi:10.1038/srep20689} {\bibfield  {journal}
  {\bibinfo  {journal} {Scientific Reports volume}\ }\textbf {\bibinfo {volume}
  {6}},\ \bibinfo {pages} {20689} (\bibinfo {year} {2015})}\BibitemShut
  {NoStop}%
\bibitem [{\citenamefont {Pedersen}\ \emph {et~al.}(2018)\citenamefont
  {Pedersen}, \citenamefont {Sch{\o}der},\ and\ \citenamefont {Dyre}}]{ped18}%
  \BibitemOpen
  \bibfield  {author} {\bibinfo {author} {\bibfnamefont {U.~R.}\ \bibnamefont
  {Pedersen}}, \bibinfo {author} {\bibfnamefont {T.~B.}\ \bibnamefont
  {Sch{\o}der}}, \ and\ \bibinfo {author} {\bibfnamefont {J.~C.}\ \bibnamefont
  {Dyre}},\ }\href {\doibase 10.1103/physrevlett.120.165501} {\bibfield
  {journal} {\bibinfo  {journal} {Phys. Rev. Lett.}\ }\textbf {\bibinfo
  {volume} {120}},\ \bibinfo {pages} {165501} (\bibinfo {year}
  {2018})}\BibitemShut {NoStop}%
\bibitem [{\citenamefont {Alba-Simionesco}\ \emph {et~al.}(2002)\citenamefont
  {Alba-Simionesco}, \citenamefont {Kivelson},\ and\ \citenamefont
  {Tarjus}}]{alb02}%
  \BibitemOpen
  \bibfield  {author} {\bibinfo {author} {\bibfnamefont {C.}~\bibnamefont
  {Alba-Simionesco}}, \bibinfo {author} {\bibfnamefont {D.}~\bibnamefont
  {Kivelson}}, \ and\ \bibinfo {author} {\bibfnamefont {G.}~\bibnamefont
  {Tarjus}},\ }\href {\doibase 10.1063/1.1452724} {\bibfield  {journal}
  {\bibinfo  {journal} {J. Chem. Phys.}\ }\textbf {\bibinfo {volume} {116}},\
  \bibinfo {pages} {5033} (\bibinfo {year} {2002})}\BibitemShut {NoStop}%
\bibitem [{\citenamefont {Alba-Simionesco}\ \emph
  {et~al.}(2004{\natexlab{b}})\citenamefont {Alba-Simionesco}, \citenamefont
  {Cailliaux}, \citenamefont {Alegria},\ and\ \citenamefont {Tarjus}}]{alb04}%
  \BibitemOpen
  \bibfield  {author} {\bibinfo {author} {\bibfnamefont {C.}~\bibnamefont
  {Alba-Simionesco}}, \bibinfo {author} {\bibfnamefont {A.}~\bibnamefont
  {Cailliaux}}, \bibinfo {author} {\bibfnamefont {A.}~\bibnamefont {Alegria}},
  \ and\ \bibinfo {author} {\bibfnamefont {G.}~\bibnamefont {Tarjus}},\
  }\href@noop {} {\bibfield  {journal} {\bibinfo  {journal} {Europhys. Lett.}\
  }\textbf {\bibinfo {volume} {68}},\ \bibinfo {pages} {58} (\bibinfo {year}
  {2004}{\natexlab{b}})}\BibitemShut {NoStop}%
\bibitem [{\citenamefont {Alba-Simionesco}\ and\ \citenamefont
  {Tarjus}(2006)}]{alb06}%
  \BibitemOpen
  \bibfield  {author} {\bibinfo {author} {\bibfnamefont {C.}~\bibnamefont
  {Alba-Simionesco}}\ and\ \bibinfo {author} {\bibfnamefont {G.}~\bibnamefont
  {Tarjus}},\ }\href@noop {} {\bibfield  {journal} {\bibinfo  {journal} {J.
  Non-Cryst. Solids}\ }\textbf {\bibinfo {volume} {352}},\ \bibinfo {pages}
  {4888} (\bibinfo {year} {2006})}\BibitemShut {NoStop}%
\bibitem [{\citenamefont {Hoover}\ \emph {et~al.}(1971)\citenamefont {Hoover},
  \citenamefont {Gray},\ and\ \citenamefont {Johnson}}]{hoo71}%
  \BibitemOpen
  \bibfield  {author} {\bibinfo {author} {\bibfnamefont {W.~G.}\ \bibnamefont
  {Hoover}}, \bibinfo {author} {\bibfnamefont {S.~G.}\ \bibnamefont {Gray}}, \
  and\ \bibinfo {author} {\bibfnamefont {K.~W.}\ \bibnamefont {Johnson}},\
  }\href@noop {} {\bibfield  {journal} {\bibinfo  {journal} {J. Chem. Phys.}\
  }\textbf {\bibinfo {volume} {55}},\ \bibinfo {pages} {1128} (\bibinfo {year}
  {1971})}\BibitemShut {NoStop}%
\bibitem [{\citenamefont {Koperwas}\ \emph {et~al.}(2012)\citenamefont
  {Koperwas}, \citenamefont {Grzybowski}, \citenamefont {Grzybowska},
  \citenamefont {Wojnarowska}, \citenamefont {Pionteck}, \citenamefont
  {Sokolov},\ and\ \citenamefont {Paluch}}]{Koperwas}%
  \BibitemOpen
  \bibfield  {author} {\bibinfo {author} {\bibfnamefont {K.}~\bibnamefont
  {Koperwas}}, \bibinfo {author} {\bibfnamefont {A.}~\bibnamefont
  {Grzybowski}}, \bibinfo {author} {\bibfnamefont {K.}~\bibnamefont
  {Grzybowska}}, \bibinfo {author} {\bibfnamefont {Z.}~\bibnamefont
  {Wojnarowska}}, \bibinfo {author} {\bibfnamefont {J.}~\bibnamefont
  {Pionteck}}, \bibinfo {author} {\bibfnamefont {A.~P.}\ \bibnamefont
  {Sokolov}}, \ and\ \bibinfo {author} {\bibfnamefont {M.}~\bibnamefont
  {Paluch}},\ }\href {\doibase 10.1103/PhysRevE.86.041502} {\bibfield
  {journal} {\bibinfo  {journal} {Phys. Rev. E}\ }\textbf {\bibinfo {volume}
  {86}},\ \bibinfo {pages} {041502} (\bibinfo {year} {2012})}\BibitemShut
  {NoStop}%
\bibitem [{\citenamefont {Papathanassiou}(2009)}]{pap09}%
  \BibitemOpen
  \bibfield  {author} {\bibinfo {author} {\bibfnamefont {A.~N.}\ \bibnamefont
  {Papathanassiou}},\ }\href@noop {} {\bibfield  {journal} {\bibinfo  {journal}
  {Phys. Rev. E}\ }\textbf {\bibinfo {volume} {79}},\ \bibinfo {pages} {032501}
  (\bibinfo {year} {2009})}\BibitemShut {NoStop}%
\bibitem [{\citenamefont {Paluch}\ \emph {et~al.}(2010)\citenamefont {Paluch},
  \citenamefont {Haracz}, \citenamefont {Grzybowski}, \citenamefont {Mierzwa},
  \citenamefont {Pionteck}, \citenamefont {Rivera-Calzada},\ and\ \citenamefont
  {Leon}}]{pal10}%
  \BibitemOpen
  \bibfield  {author} {\bibinfo {author} {\bibfnamefont {M.}~\bibnamefont
  {Paluch}}, \bibinfo {author} {\bibfnamefont {S.}~\bibnamefont {Haracz}},
  \bibinfo {author} {\bibfnamefont {A.}~\bibnamefont {Grzybowski}}, \bibinfo
  {author} {\bibfnamefont {M.}~\bibnamefont {Mierzwa}}, \bibinfo {author}
  {\bibfnamefont {J.}~\bibnamefont {Pionteck}}, \bibinfo {author}
  {\bibfnamefont {A.}~\bibnamefont {Rivera-Calzada}}, \ and\ \bibinfo {author}
  {\bibfnamefont {C.}~\bibnamefont {Leon}},\ }\href@noop {} {\bibfield
  {journal} {\bibinfo  {journal} {J. Phys. Chem. Lett.}\ }\textbf {\bibinfo
  {volume} {1}},\ \bibinfo {pages} {987} (\bibinfo {year} {2010})}\BibitemShut
  {NoStop}%
\bibitem [{\citenamefont {Romanini}\ \emph {et~al.}(2017)\citenamefont
  {Romanini}, \citenamefont {Barrio}, \citenamefont {Macovez}, \citenamefont
  {Ruiz-Martin}, \citenamefont {Capaccioli},\ and\ \citenamefont
  {Tamarit}}]{Romanini}%
  \BibitemOpen
  \bibfield  {author} {\bibinfo {author} {\bibfnamefont {M.}~\bibnamefont
  {Romanini}}, \bibinfo {author} {\bibfnamefont {M.}~\bibnamefont {Barrio}},
  \bibinfo {author} {\bibfnamefont {R.}~\bibnamefont {Macovez}}, \bibinfo
  {author} {\bibfnamefont {M.~D.}\ \bibnamefont {Ruiz-Martin}}, \bibinfo
  {author} {\bibfnamefont {S.}~\bibnamefont {Capaccioli}}, \ and\ \bibinfo
  {author} {\bibfnamefont {J.~L.}\ \bibnamefont {Tamarit}},\ }\href {\doibase
  10.1038/s41598-017-01464-2} {\bibfield  {journal} {\bibinfo  {journal}
  {Scientific Reports}\ }\textbf {\bibinfo {volume} {7}} (\bibinfo {year}
  {2017}),\ 10.1038/s41598-017-01464-2}\BibitemShut {NoStop}%
\bibitem [{\citenamefont {Angell}(1985)}]{ang85}%
  \BibitemOpen
  \bibfield  {author} {\bibinfo {author} {\bibfnamefont {C.~A.}\ \bibnamefont
  {Angell}},\ }\enquote {\bibinfo {title} {Relaxations in complex systems},}\ \
  (\bibinfo  {publisher} {NRL, Washington,},\ \bibinfo {year}
  {1985})\BibitemShut {NoStop}%
\bibitem [{\citenamefont {Angell}(1995)}]{Angell2}%
  \BibitemOpen
  \bibfield  {author} {\bibinfo {author} {\bibfnamefont {C.~A.}\ \bibnamefont
  {Angell}},\ }\href {http://www.jstor.org/stable/2886440} {\bibfield
  {journal} {\bibinfo  {journal} {Science}\ }\textbf {\bibinfo {volume}
  {267}},\ \bibinfo {pages} {1924} (\bibinfo {year} {1995})}\BibitemShut
  {NoStop}%
\bibitem [{\citenamefont {Tarjus}\ and\ \citenamefont
  {Alba-Simionesco}(2014)}]{tar14}%
  \BibitemOpen
  \bibfield  {author} {\bibinfo {author} {\bibfnamefont {G.}~\bibnamefont
  {Tarjus}}\ and\ \bibinfo {author} {\bibfnamefont {C.}~\bibnamefont
  {Alba-Simionesco}},\ }\enquote {\bibinfo {title} {Fragility of glass-forming
  liquids},}\ \ (\bibinfo  {publisher} {Hindustan Book Agency},\ \bibinfo
  {year} {2014})\ Chap.\ \bibinfo {chapter} {An assessment of the concept of
  fragility}\BibitemShut {NoStop}%
\bibitem [{\citenamefont {Angell}(1988)}]{Angell}%
  \BibitemOpen
  \bibfield  {author} {\bibinfo {author} {\bibfnamefont {C.~A.}\ \bibnamefont
  {Angell}},\ }\href {\doibase 10.1016/0022-3093(88)90133-0} {\bibfield
  {journal} {\bibinfo  {journal} {Journal of Non-Crystalline Solids}\ }\textbf
  {\bibinfo {volume} {102}},\ \bibinfo {pages} {205} (\bibinfo {year}
  {1988})}\BibitemShut {NoStop}%
\bibitem [{\citenamefont {Christensen}\ and\ \citenamefont
  {Olsen}(1994)}]{Tage}%
  \BibitemOpen
  \bibfield  {author} {\bibinfo {author} {\bibfnamefont {T.}~\bibnamefont
  {Christensen}}\ and\ \bibinfo {author} {\bibfnamefont {N.~B.}\ \bibnamefont
  {Olsen}},\ }\href {\doibase 10.1103/PhysRevB.49.15396} {\bibfield  {journal}
  {\bibinfo  {journal} {Phys. Rev. B}\ }\textbf {\bibinfo {volume} {49}},\
  \bibinfo {pages} {15396} (\bibinfo {year} {1994})}\BibitemShut {NoStop}%
\bibitem [{\citenamefont {Klein}\ and\ \citenamefont {Angell}(2016)}]{Klein}%
  \BibitemOpen
  \bibfield  {author} {\bibinfo {author} {\bibfnamefont {I.~S.}\ \bibnamefont
  {Klein}}\ and\ \bibinfo {author} {\bibfnamefont {C.~A.}\ \bibnamefont
  {Angell}},\ }\href@noop {} {\bibfield  {journal} {\bibinfo  {journal}
  {Journal of Non-Crystalline Solids}\ }\textbf {\bibinfo {volume} {451}},\
  \bibinfo {pages} {116} (\bibinfo {year} {2016})}\BibitemShut {NoStop}%
\bibitem [{\citenamefont {Roed}\ \emph {et~al.}(2013)\citenamefont {Roed},
  \citenamefont {Gundermann}, \citenamefont {Dyre},\ and\ \citenamefont
  {Niss}}]{Lisa}%
  \BibitemOpen
  \bibfield  {author} {\bibinfo {author} {\bibfnamefont {L.~A.}\ \bibnamefont
  {Roed}}, \bibinfo {author} {\bibfnamefont {D.}~\bibnamefont {Gundermann}},
  \bibinfo {author} {\bibfnamefont {J.~C.}\ \bibnamefont {Dyre}}, \ and\
  \bibinfo {author} {\bibfnamefont {K.}~\bibnamefont {Niss}},\ }\href@noop {}
  {\bibfield  {journal} {\bibinfo  {journal} {Journal of Chemical Physics}\
  }\textbf {\bibinfo {volume} {139}},\ \bibinfo {pages} {101101} (\bibinfo
  {year} {2013})}\BibitemShut {NoStop}%
\bibitem [{\citenamefont {Casalini}\ \emph {et~al.}(2011)\citenamefont
  {Casalini}, \citenamefont {Gamache},\ and\ \citenamefont
  {Roland}}]{Casalini3}%
  \BibitemOpen
  \bibfield  {author} {\bibinfo {author} {\bibfnamefont {R.}~\bibnamefont
  {Casalini}}, \bibinfo {author} {\bibfnamefont {R.~F.}\ \bibnamefont
  {Gamache}}, \ and\ \bibinfo {author} {\bibfnamefont {C.~M.}\ \bibnamefont
  {Roland}},\ }\href {\doibase 10.1063/1.3664180} {\bibfield  {journal}
  {\bibinfo  {journal} {The Journal of Chemical Physics}\ }\textbf {\bibinfo
  {volume} {135}},\ \bibinfo {pages} {224501} (\bibinfo {year}
  {2011})}\BibitemShut {NoStop}%
\bibitem [{\citenamefont {Gundermann}(2013)}]{Ditte2}%
  \BibitemOpen
  \bibfield  {author} {\bibinfo {author} {\bibfnamefont {D.}~\bibnamefont
  {Gundermann}},\ }\emph {\bibinfo {title} {Testing Predictions of the Isomorph
  Theory by Experiment}},\ \href
  {http://glass.ruc.dk/pdf/phd_afhandlinger/ditte_thesis.pdf} {Ph.D. thesis},\
  \bibinfo  {school} {IMFUFA, Roskilde University} (\bibinfo {year}
  {2013})\BibitemShut {NoStop}%
\bibitem [{\citenamefont {Adrjanowicz}\ \emph {et~al.}(2017)\citenamefont
  {Adrjanowicz}, \citenamefont {Kaminski}, \citenamefont {Tarnacka},
  \citenamefont {Szklarz},\ and\ \citenamefont {Paluch}}]{Adrjanowicz}%
  \BibitemOpen
  \bibfield  {author} {\bibinfo {author} {\bibfnamefont {K.}~\bibnamefont
  {Adrjanowicz}}, \bibinfo {author} {\bibfnamefont {K.}~\bibnamefont
  {Kaminski}}, \bibinfo {author} {\bibfnamefont {M.}~\bibnamefont {Tarnacka}},
  \bibinfo {author} {\bibfnamefont {G.}~\bibnamefont {Szklarz}}, \ and\
  \bibinfo {author} {\bibfnamefont {M.}~\bibnamefont {Paluch}},\ }\href@noop {}
  {\bibfield  {journal} {\bibinfo  {journal} {The Journal of Physical Chemistry
  Letters}\ }\textbf {\bibinfo {volume} {8}},\ \bibinfo {pages} {696} (\bibinfo
  {year} {2017})}\BibitemShut {NoStop}%
\bibitem [{\citenamefont {Dreyfus}\ \emph {et~al.}(2004)\citenamefont
  {Dreyfus}, \citenamefont {Le~Grand}, \citenamefont {Gapinski}, \citenamefont
  {Steffen},\ and\ \citenamefont {Patkowski}}]{Dreyfus}%
  \BibitemOpen
  \bibfield  {author} {\bibinfo {author} {\bibfnamefont {C.}~\bibnamefont
  {Dreyfus}}, \bibinfo {author} {\bibfnamefont {A.}~\bibnamefont {Le~Grand}},
  \bibinfo {author} {\bibfnamefont {J.}~\bibnamefont {Gapinski}}, \bibinfo
  {author} {\bibfnamefont {W.}~\bibnamefont {Steffen}}, \ and\ \bibinfo
  {author} {\bibfnamefont {A.}~\bibnamefont {Patkowski}},\ }\href {\doibase
  10.1140/epjb/e2004-00386-3} {\bibfield  {journal} {\bibinfo  {journal} {Eur.
  Phys. J. B}\ }\textbf {\bibinfo {volume} {42}},\ \bibinfo {pages} {309}
  (\bibinfo {year} {2004})}\BibitemShut {NoStop}%
\bibitem [{\citenamefont {Klieber}\ \emph {et~al.}(2013)\citenamefont
  {Klieber}, \citenamefont {Hecksher}, \citenamefont {Pezeril}, \citenamefont
  {Torchinsky}, \citenamefont {Dyre},\ and\ \citenamefont {Nelson}}]{Klieber}%
  \BibitemOpen
  \bibfield  {author} {\bibinfo {author} {\bibfnamefont {C.}~\bibnamefont
  {Klieber}}, \bibinfo {author} {\bibfnamefont {T.}~\bibnamefont {Hecksher}},
  \bibinfo {author} {\bibfnamefont {T.}~\bibnamefont {Pezeril}}, \bibinfo
  {author} {\bibfnamefont {D.~H.}\ \bibnamefont {Torchinsky}}, \bibinfo
  {author} {\bibfnamefont {J.~C.}\ \bibnamefont {Dyre}}, \ and\ \bibinfo
  {author} {\bibfnamefont {K.~A.}\ \bibnamefont {Nelson}},\ }\href {\doibase
  10.1063/1.4789948} {\bibfield  {journal} {\bibinfo  {journal} {The Journal of
  Chemical Physics}\ }\textbf {\bibinfo {volume} {138}},\ \bibinfo {pages}
  {12A544} (\bibinfo {year} {2013})}\BibitemShut {NoStop}%
\bibitem [{\citenamefont {Roland}\ \emph {et~al.}(2005)\citenamefont {Roland},
  \citenamefont {Hensel-Bielowka}, \citenamefont {Paluch},\ and\ \citenamefont
  {Casalini}}]{Roland}%
  \BibitemOpen
  \bibfield  {author} {\bibinfo {author} {\bibfnamefont {C.~M.}\ \bibnamefont
  {Roland}}, \bibinfo {author} {\bibfnamefont {S.}~\bibnamefont
  {Hensel-Bielowka}}, \bibinfo {author} {\bibfnamefont {M.}~\bibnamefont
  {Paluch}}, \ and\ \bibinfo {author} {\bibfnamefont {R.}~\bibnamefont
  {Casalini}},\ }\href@noop {} {\bibfield  {journal} {\bibinfo  {journal}
  {Reports on Progress in Physics}\ }\textbf {\bibinfo {volume} {68}},\
  \bibinfo {pages} {1405} (\bibinfo {year} {2005})}\BibitemShut {NoStop}%
\bibitem [{\citenamefont {Grzybowski}\ \emph {et~al.}(2006)\citenamefont
  {Grzybowski}, \citenamefont {Grzybowska}, \citenamefont {Zio\l{}o},\ and\
  \citenamefont {Paluch}}]{Grzybowski2}%
  \BibitemOpen
  \bibfield  {author} {\bibinfo {author} {\bibfnamefont {A.}~\bibnamefont
  {Grzybowski}}, \bibinfo {author} {\bibfnamefont {K.}~\bibnamefont
  {Grzybowska}}, \bibinfo {author} {\bibfnamefont {J.}~\bibnamefont
  {Zio\l{}o}}, \ and\ \bibinfo {author} {\bibfnamefont {M.}~\bibnamefont
  {Paluch}},\ }\href {\doibase 10.1103/PhysRevE.74.041503} {\bibfield
  {journal} {\bibinfo  {journal} {Phys. Rev. E}\ }\textbf {\bibinfo {volume}
  {74}},\ \bibinfo {pages} {041503} (\bibinfo {year} {2006})}\BibitemShut
  {NoStop}%
\bibitem [{\citenamefont {Sanz}(lane)}]{Preparation}%
  \BibitemOpen
  \bibfield  {author} {\bibinfo {author} {\bibfnamefont {A.}~\bibnamefont
  {Sanz}},\ }\href@noop {} {\enquote {\bibinfo {title} {Unpublished results},}\
  } (\bibinfo {year} {Density scaling in dipropylene glycol by dielectric
  spectroscopy measurements in pressure-temperature plane})\BibitemShut
  {NoStop}%
\bibitem [{\citenamefont {Bailey}\ \emph {et~al.}(2017)\citenamefont {Bailey},
  \citenamefont {Ingebrigtsen}, \citenamefont {Hansen}, \citenamefont
  {Veldhorst}, \citenamefont {Bøhling}, \citenamefont {Lemarchand},
  \citenamefont {Olsen}, \citenamefont {Bacher}, \citenamefont {Costigliola},
  \citenamefont {Pedersen}, \citenamefont {Larsen}, \citenamefont {Dyre},\ and\
  \citenamefont {Schrøder}}]{rumd}%
  \BibitemOpen
  \bibfield  {author} {\bibinfo {author} {\bibfnamefont {N.~P.}\ \bibnamefont
  {Bailey}}, \bibinfo {author} {\bibfnamefont {T.~S.}\ \bibnamefont
  {Ingebrigtsen}}, \bibinfo {author} {\bibfnamefont {J.~S.}\ \bibnamefont
  {Hansen}}, \bibinfo {author} {\bibfnamefont {A.~A.}\ \bibnamefont
  {Veldhorst}}, \bibinfo {author} {\bibfnamefont {L.}~\bibnamefont {Bøhling}},
  \bibinfo {author} {\bibfnamefont {C.~A.}\ \bibnamefont {Lemarchand}},
  \bibinfo {author} {\bibfnamefont {A.~E.}\ \bibnamefont {Olsen}}, \bibinfo
  {author} {\bibfnamefont {A.~K.}\ \bibnamefont {Bacher}}, \bibinfo {author}
  {\bibfnamefont {L.}~\bibnamefont {Costigliola}}, \bibinfo {author}
  {\bibfnamefont {U.~R.}\ \bibnamefont {Pedersen}}, \bibinfo {author}
  {\bibfnamefont {H.}~\bibnamefont {Larsen}}, \bibinfo {author} {\bibfnamefont
  {J.~C.}\ \bibnamefont {Dyre}}, \ and\ \bibinfo {author} {\bibfnamefont
  {T.~B.}\ \bibnamefont {Schrøder}},\ }\href@noop {} {\bibfield  {journal}
  {\bibinfo  {journal} {SciPost Physics}\ } (\bibinfo {year}
  {2017})}\BibitemShut {NoStop}%
\bibitem [{\citenamefont {Lennard-Jones}(1924)}]{lj24}%
  \BibitemOpen
  \bibfield  {author} {\bibinfo {author} {\bibfnamefont {J.~E.}\ \bibnamefont
  {Lennard-Jones}},\ }\href@noop {} {\bibfield  {journal} {\bibinfo  {journal}
  {Proc. R. Soc. London A}\ }\textbf {\bibinfo {volume} {106}},\ \bibinfo
  {pages} {441} (\bibinfo {year} {1924})}\BibitemShut {NoStop}%
\bibitem [{\citenamefont {Lewis}\ and\ \citenamefont
  {Wahnstr{\"o}m}(1993)}]{lew93}%
  \BibitemOpen
  \bibfield  {author} {\bibinfo {author} {\bibfnamefont {L.~J.}\ \bibnamefont
  {Lewis}}\ and\ \bibinfo {author} {\bibfnamefont {G.}~\bibnamefont
  {Wahnstr{\"o}m}},\ }\href@noop {} {\bibfield  {journal} {\bibinfo  {journal}
  {Solid State Commun}\ }\textbf {\bibinfo {volume} {86}},\ \bibinfo {pages}
  {295} (\bibinfo {year} {1993})}\BibitemShut {NoStop}%
\bibitem [{\citenamefont {Singh}\ \emph {et~al.}(2007)\citenamefont {Singh},
  \citenamefont {Arafin},\ and\ \citenamefont {George}}]{sin07}%
  \BibitemOpen
  \bibfield  {author} {\bibinfo {author} {\bibfnamefont {R.~N.}\ \bibnamefont
  {Singh}}, \bibinfo {author} {\bibfnamefont {S.}~\bibnamefont {Arafin}}, \
  and\ \bibinfo {author} {\bibfnamefont {A.~K.}\ \bibnamefont {George}},\
  }\href@noop {} {\bibfield  {journal} {\bibinfo  {journal} {Physica B}\
  }\textbf {\bibinfo {volume} {387}},\ \bibinfo {pages} {344} (\bibinfo {year}
  {2007})}\BibitemShut {NoStop}%
\bibitem [{\citenamefont {Hummel}\ \emph {et~al.}(2015)\citenamefont {Hummel},
  \citenamefont {Kresse}, \citenamefont {Dyre},\ and\ \citenamefont
  {Pedersen}}]{hum15}%
  \BibitemOpen
  \bibfield  {author} {\bibinfo {author} {\bibfnamefont {F.}~\bibnamefont
  {Hummel}}, \bibinfo {author} {\bibfnamefont {G.}~\bibnamefont {Kresse}},
  \bibinfo {author} {\bibfnamefont {J.~C.}\ \bibnamefont {Dyre}}, \ and\
  \bibinfo {author} {\bibfnamefont {U.~R.}\ \bibnamefont {Pedersen}},\ }\href
  {\doibase 10.1103/PhysRevB.92.174116} {\bibfield  {journal} {\bibinfo
  {journal} {Phys. Rev. B}\ }\textbf {\bibinfo {volume} {92}},\ \bibinfo
  {pages} {174116} (\bibinfo {year} {2015})}\BibitemShut {NoStop}%
\bibitem [{\citenamefont {Gr{\"u}neisen}(1912)}]{gru12}%
  \BibitemOpen
  \bibfield  {author} {\bibinfo {author} {\bibfnamefont {E.}~\bibnamefont
  {Gr{\"u}neisen}},\ }\href {\doibase 10.1002/andp.19123441202} {\bibfield
  {journal} {\bibinfo  {journal} {Ann. Physik}\ }\textbf {\bibinfo {volume}
  {12}},\ \bibinfo {pages} {257} (\bibinfo {year} {1912})}\BibitemShut
  {NoStop}%
\bibitem [{\citenamefont {B{\o}hling}\ \emph {et~al.}(2012)\citenamefont
  {B{\o}hling}, \citenamefont {Ingebrigtsen}, \citenamefont {Grzybowski},
  \citenamefont {Paluch}, \citenamefont {Dyre},\ and\ \citenamefont
  {Sch{\o}der}}]{Bohling}%
  \BibitemOpen
  \bibfield  {author} {\bibinfo {author} {\bibfnamefont {L.}~\bibnamefont
  {B{\o}hling}}, \bibinfo {author} {\bibfnamefont {T.~S.}\ \bibnamefont
  {Ingebrigtsen}}, \bibinfo {author} {\bibfnamefont {A.}~\bibnamefont
  {Grzybowski}}, \bibinfo {author} {\bibfnamefont {M.}~\bibnamefont {Paluch}},
  \bibinfo {author} {\bibfnamefont {J.~C.}\ \bibnamefont {Dyre}}, \ and\
  \bibinfo {author} {\bibfnamefont {T.~B.}\ \bibnamefont {Sch{\o}der}},\ }\href
  {http://stacks.iop.org/1367-2630/14/i=11/a=113035} {\bibfield  {journal}
  {\bibinfo  {journal} {New Journal of Physics}\ }\textbf {\bibinfo {volume}
  {14}},\ \bibinfo {pages} {113035} (\bibinfo {year} {2012})}\BibitemShut
  {NoStop}%
\bibitem [{\citenamefont {Grzybowski}\ \emph {et~al.}(2014)\citenamefont
  {Grzybowski}, \citenamefont {Koperwas},\ and\ \citenamefont
  {Paluch}}]{Grzybowski}%
  \BibitemOpen
  \bibfield  {author} {\bibinfo {author} {\bibfnamefont {A.}~\bibnamefont
  {Grzybowski}}, \bibinfo {author} {\bibfnamefont {K.}~\bibnamefont
  {Koperwas}}, \ and\ \bibinfo {author} {\bibfnamefont {M.}~\bibnamefont
  {Paluch}},\ }\href {\doibase 10.1063/1.4861907} {\bibfield  {journal}
  {\bibinfo  {journal} {The Journal of Chemical Physics}\ }\textbf {\bibinfo
  {volume} {140}},\ \bibinfo {pages} {044502} (\bibinfo {year}
  {2014})}\BibitemShut {NoStop}%
\bibitem [{\citenamefont {L\'opez}\ \emph {et~al.}(2018)\citenamefont
  {L\'opez}, \citenamefont {Fandino}, \citenamefont {Cabaleiro}, \citenamefont
  {Lugo},\ and\ \citenamefont {Fern\'andez}}]{Lopez}%
  \BibitemOpen
  \bibfield  {author} {\bibinfo {author} {\bibfnamefont {E.~R.}\ \bibnamefont
  {L\'opez}}, \bibinfo {author} {\bibfnamefont {O.}~\bibnamefont {Fandino}},
  \bibinfo {author} {\bibfnamefont {D.}~\bibnamefont {Cabaleiro}}, \bibinfo
  {author} {\bibfnamefont {L.}~\bibnamefont {Lugo}}, \ and\ \bibinfo {author}
  {\bibfnamefont {J.}~\bibnamefont {Fern\'andez}},\ }\href {\doibase
  10.1039/C7CP07180A} {\bibfield  {journal} {\bibinfo  {journal} {Phys. Chem.
  Chem. Phys.}\ }\textbf {\bibinfo {volume} {20}},\ \bibinfo {pages} {3531}
  (\bibinfo {year} {2018})}\BibitemShut {NoStop}%
\end{thebibliography}%

\end{document}